\documentclass{appolb}
\usepackage{graphicx}

\usepackage{amssymb}
\usepackage{amsmath}
\usepackage{float}

\begin{document}

\title{DESIGN AND PRODUCTION OF A FREE-FORM LENS FOR LONG RANGE LED ILLUMINATION}
\author{Ahmet Bing\"ul, Mehmet Adiyaman
\address{Department of Engineering of Physics, Gaziantep University, Gaziantep 27310, Turkey.}
\\
}
\maketitle

\begin{abstract}
In this study, a procedure for designing a free-form lens
for long-range LED illumination is presented.
The geometrical form of the proposed lens is obtained by minimizing 
optical path lengths of the rays emitted from a point-like light source.
Optical ray tracing simulations of two different LEDs and the free-form 
lens are performed by using Zemax OpticStudio.
In addition, the prototype of the free-form lens is manufactured 
by the plastic injection molding method using PMMA material.
Nine of the lenses are used to build an LED projector 
in the form of a 3x3 lens matrix. 
The optical measurements of the projector are compared with the 
results predicted in the simulations. It is found that the beam
divergence of the projector is less than 10 degrees when using 
suitable LEDs in visible and near infrared regions.
\end{abstract}

\section{Introduction}
LED (Light Emitting Diode) is an energy-saving source providing 
high light efficiency and has a long life time. 
Despite the high light efficiency, in general, 
the direct light output of LEDs spread at wide angle. 
This results in a disadvantage in the use of LEDs when illuminating a 
surface being at a large distance from the source. The light intensity 
on a target can be improved either by using reflectors or 
by placing the narrow-angle free-form lenses having a suitable geometry 
in front of the LEDs.

There are many academic and engineering studies 
on design and production of free-form lenses in literature
such as~\cite{Ries, Ding, Yang, Anh, Hu, Lin}. 
These studies are mostly based on defining a 
differential equation related to design 
and its numerical solution to obtain the geometrical form of the lens.
In this study, however, a free-form lens design based on 
optical path length calculations is presented for long-range (or narrow-angle) 
illumination for both near infrared (IR) and visible regions.
Optical ray tracing analyses of LEDs and the free-form lens are developed 
in Zemax OpticStudio 20.1~\cite{Zemax}. Three different polymers are used as 
lens materials; PMMA (poly(methyl methacrylate)), PC (polycarbonate) and PS (polystyrene).
Simulation studies show that PMMA exhibits a better performance
in terms of total optical power on the detector compared to others.

In addition, the prototype of solid form of the lens made from PMMA 
is manufactured via plastic injection molding method. 
Nine of them are used to build a projector which is basically a 3x3 lens array.
By using two different LEDs (white and IR), optical measurements of the 
projector are compared with the predictions of Zemax.
Results seem to be consistent with that of simulations and
the beam divergence of the projector for visible and IR regions
are found to be less than $10^\circ$.

\section{Free-Form Lens Profile}
Free-form lenses are preferred in both illumination~\cite{Ding, Lin} and 
imaging systems~\cite{Yang}. In this study, a profile of a free-form lens to be used in LED illumination 
is proposed in order to obtain a narrow-angle beam spot at long distances.

\begin{figure}[H]
	\centering
	\includegraphics[width=0.80\textwidth]{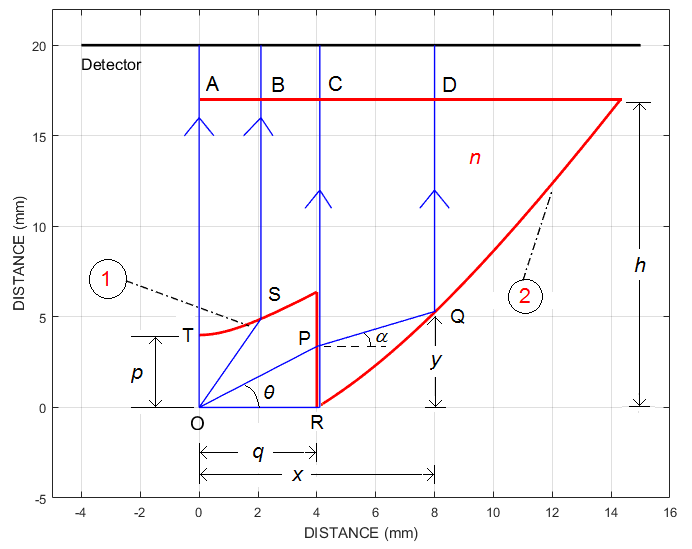}
	\caption{Surfaces of free-form lens and rays used in the analysis.}
	\label{fig_lens}
\end{figure}

The free-form lens (or simply referred to as lens) is assumed to be used in air and has 
index of refraction $n$. For simplicity, one can start the design with a point-like 
source placed at the origin of x-y coordinate system and with one-dimensional curves 
representing the \textit{surfaces} of the lens.
Aim of the lens is to deflect all of the 
rays emitted from the source such that they are parallel to 
optical axis on the exit side of the lens. 
Figure~\ref{fig_lens} shows the selected curves and 
four rays originating from the point source placed at origin.
Shape of the surface 1 and 2 are non-linear functions. The remaining
surfaces are represented by lines.
Therefore, there are four design parameters which affect the 
optical performance and geometrical size of the lens; 
the lengths $p$, $q$, $h$ and refractive index $n$.

The shape of the first surface can be found as follows.
We desire that the optical path lengths (OPL)\footnote{OPL is defined as the product of the 
	refractive index of the medium and geometrical path traversed by a ray. 
	Time of flight for the ray in the medium is given by $t = OPL/c$ where $c$ is the speed of light in vacuum.}
for rays A and B must be equal so that they reach the detector 
(or target) at the same time, namely, $OPL_A = OPL_B$ where 
$OPL_A = |OT| + n|TA|$ and $OPL_B = |OS| + n|SB|$. By substituting the lengths we obtain
\begin{equation}
\label{eqn_OPLAB}
p + n(h-p) = \sqrt{x^2+y^2} + n(h-y)
\end{equation}
where $p$ is the shortest distance between origin and surface 1, and $x \le q$.
Solving for $y$ yields two roots:
\begin{equation}
\label{eqn_hyperbola}
y(x) = \frac{np}{n+1} \pm \sqrt{ \left( \frac{x}{n-1} \right)^2 +  \left( \frac{p}{n+1} \right)^2}
\end{equation}
This is a well-known curve called the hyperbola whose focus is 
at the origin. Positive root, $y(x)>0$, is used in the analysis. 

\begin{figure}[H]
	\centering
	\includegraphics[width=0.60\textwidth]{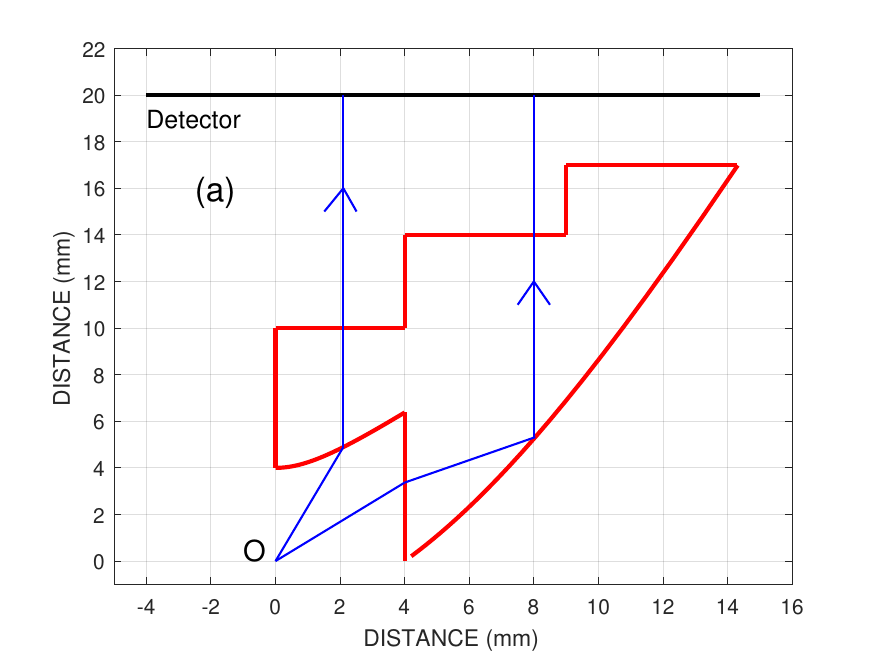}
	\includegraphics[width=0.38\textwidth]{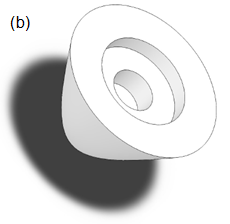}
	\caption{Schematic diagram illustrating the surfaces of final lens for $p=q=4$ mm, $h=17$ mm and $n=1.49$  
		     (a) in 2D and (b) in 3D. }
	\label{fig_2d3d}
\end{figure}

Surface 2 is slightly complicated since ray C and D will be first refracted
on the plane at point R and P respectively. Then, they are totally reflected 
at point R and Q. Again, we desire their optical paths must be the same
and they are respectively given by $OPL_C = |OR| + |RC|$ and $OPL_D = |OP| + n|PQ| + n|QD|$.
Ray D emerges from point O and makes angle $\theta$ with x-axis. 
At point P, Snell's law can be applied to extract refraction angle ($\alpha$)
from the equation $\sin\theta = n\sin\alpha$. 
By equating optical paths, we obtain
\begin{equation}
\label{eqn_OPLCD}
  q + nh = \sqrt{y_1^2+q^2} + n\sqrt{(y-y_1)^2 + (x-q)^2} + n(h-y)
\end{equation}
where $y_1$ is the distance $|PR|$ shown in Figure~\ref{fig_lens}, namely $y_1 = q\tan\theta$. 
Also, for the line passing thorough points P and Q, we have the relation 
\begin{equation}
\label{eqn_line}
y-y_1 = \tan\alpha(x-q)
\end{equation}
Frankly speaking, it is difficult to find an 
explicit relation between $x$ and $y$ as in Equation~\ref{eqn_hyperbola} 
since both $x$ and $y$ are functions of the angle $\theta$. 
However, one can obtain two parametric equations representing the surface 2.
After doing some algebra, height of the ray at surface 2 can be found as:
\begin{equation}
\label{eqn_surf2y}
y(\theta)  = \frac{q(n^2\sec\theta + 1 - \sec\theta)}{n^2 \text{cosec}\theta-1}
\end{equation}
By using Equation~\ref{eqn_line}, the abscissa can be evaluated from:
\begin{equation}
\label{eqn_surf2x}
x(\theta) = q + \frac{y - q \tan\theta}{\tan(\arcsin[\sin(\theta)/n])}
\end{equation}
In the manufacturing process, the center part of the lens is removed as a 
stairs shape to reduce weight and material cost as shown in Figure~\ref{fig_2d3d}a.
Of course, this change does not effect the illumination performance
since the top surface of the lens (before detector) is nevertheless a plane.
Finally, three-dimensional (3D) solid form of the lens is simply obtained by
rotating the lens in y-axis as shown in Figure~\ref{fig_2d3d}b.

\section{Optical Simulation}
The equations derived for 2D free-form lens geometry
and the point-like source are implemented in a Matlab programming language 
to attain initial calculations related to ray tracing for single wavelength.
However, in practice one needs 3D geometry and LED which is neither a 
point-like nor a monochromatic source.

Fortunately, there are solid modeling softwares to obtain 
3D geometry and assign desired polymeric material to the lens.
Also, LED manufacturers distribute comprehensive ray-tracing data files
to be used in optical simulations such as eulumdat, ray and spectrum files.
In principle, LED is considered to be a point source in eulumdat file
which is used for a quick analysis.
Whereas, the ray file represents actual spatial and angular 
distribution of rays originating from the outer surface of LED. 
Therefore, ray files can be used in more realistic simulations.
The spectral distribution of LED (wavelengths emitted and corresponding weights) are stored in spectrum files.

In this study, Zemax OpticStudio 20.1, which is very good at comprehensive ray tracing 
software, is used for the optical simulations. In Non-Sequential Mode of Zemax, 
one can include simulation files of LED, the 3D geometry of the lens and a 
suitable rectangular detector which is basically a counter of rays hitting on it.
PMMA $(\text{C}_{5} \text{H}_{8} \text{O}_{2})_n$ material is 
assigned to the solid model of the lens since its transmittance for the visible and near 
infrared optical regions is greater than 90\%.

\begin{figure}[H]
	\centering
	\includegraphics[width=0.49\textwidth]{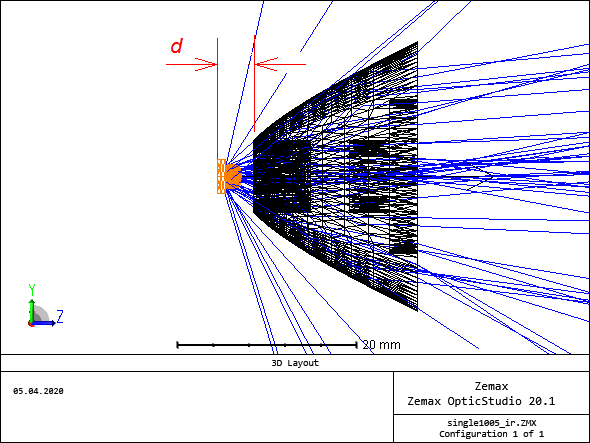}
	\includegraphics[width=0.49\textwidth]{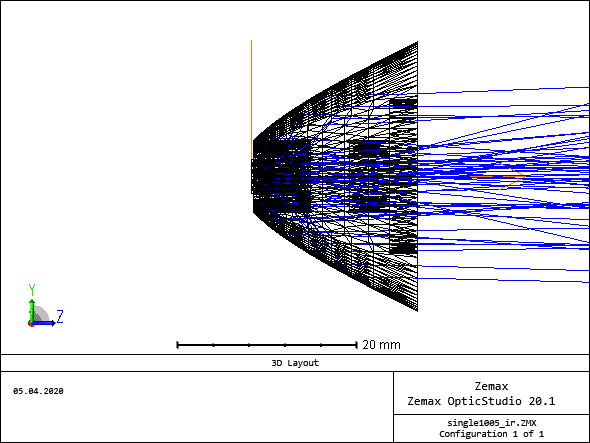}
	\caption{The 3D lens and IR LED imported to Zemax.
		The rays in layouts are obtained from the LED's ray file.
		Left, LED is placed at $d = 4$ mm back of the lens. 
		This little misalignment results in a wide angle ray distribution.
		Right, LED is placed its optimum position (at $d=0.1$ mm) where the
	    outgoing rays are almost parallel to +z-axis defined from left to right in Zemax.}
	\label{fig_simlens}
\end{figure}

Two types of LED provided by Osram Company~\cite{Osram} are 
chosen in simulations. They are SFH 4718A which is an IR LED whose peak irradiance 
is at 850 nm and LUW H9GP a \textit{white} LED having color temperature of 
6500 K. 3D lens and IR LED used in this study are shown in
Figure~\ref{fig_simlens} for two different separation, $d$, between lens and LED.
$d$ is one of the important parameter in the design 
since it significantly affects the light distribution and number of photon hits 
on the detector. In this study, $d = 0.1$ mm is found to be optimum for both LED models.

\begin{figure}[H]
	\centering
	\includegraphics[width=0.49\textwidth]{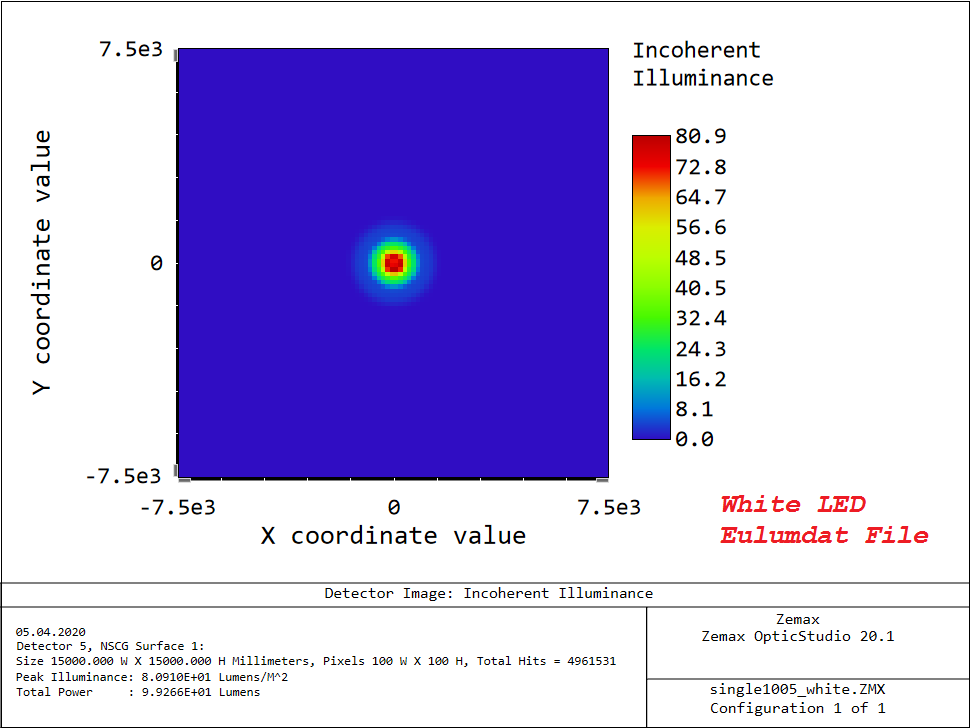}
	\includegraphics[width=0.49\textwidth]{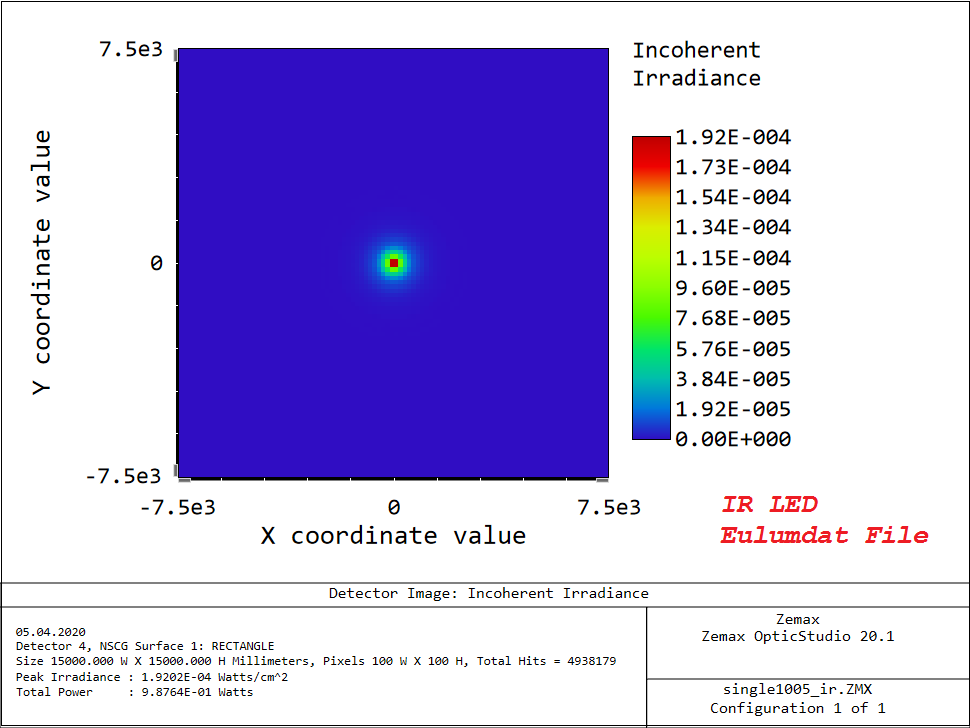}
	\includegraphics[width=0.49\textwidth]{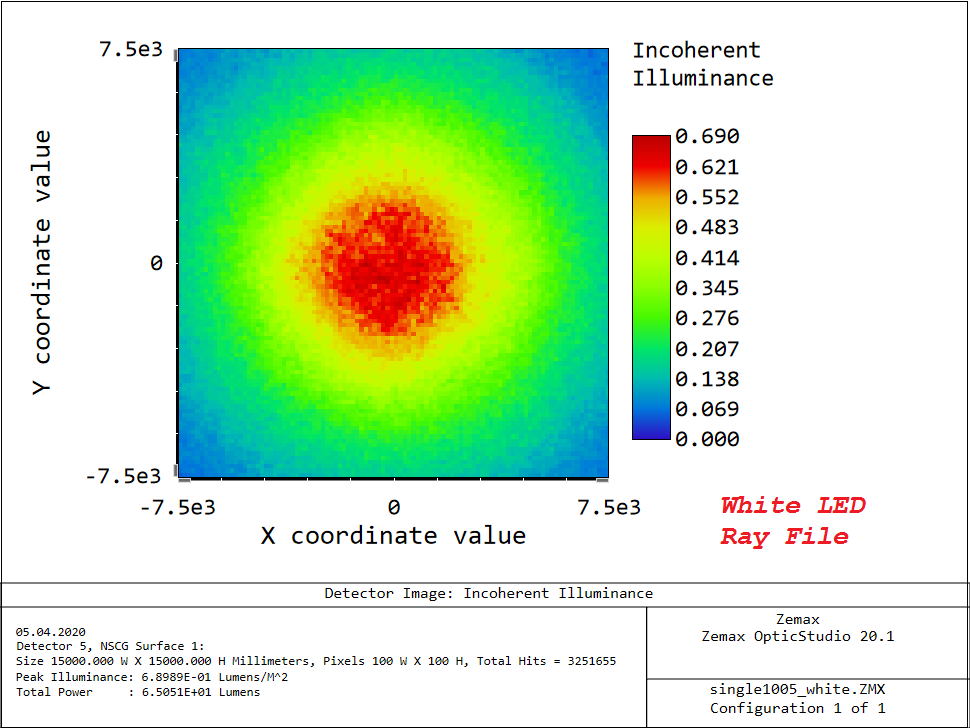}
	\includegraphics[width=0.49\textwidth]{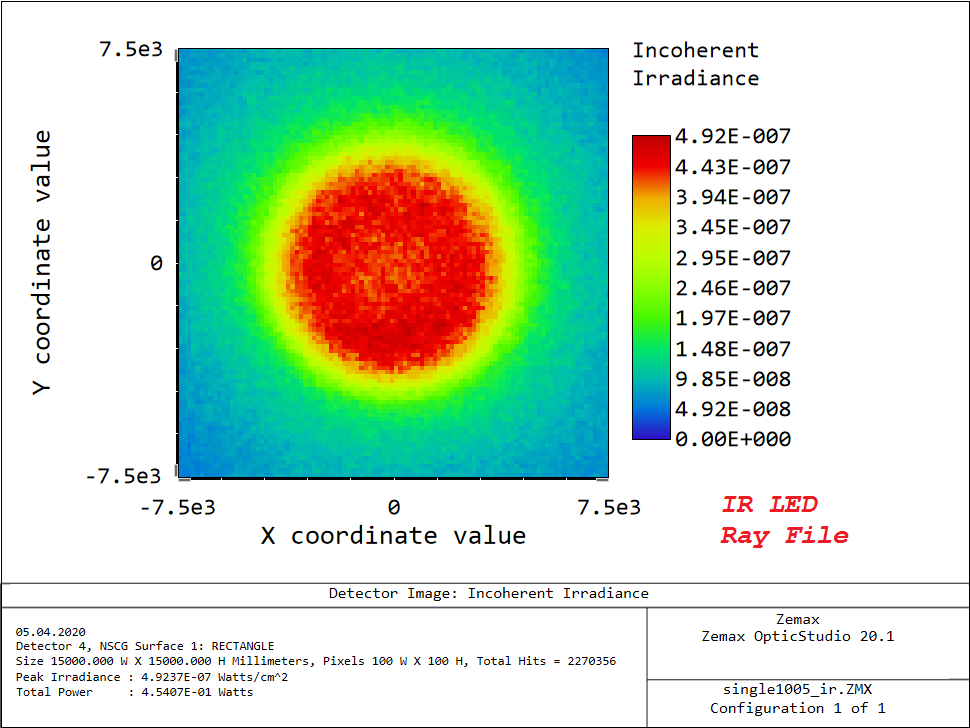}
	\caption{Spot diagrams obtained on a rectangular detector being 100 m away 
		from LED and lens system. Eulumdat, ray and spectrum files are used in ray tracing.}
	\label{fig_spot}
\end{figure}

In the analyses, totally 5 million rays (both from eulumdat file and ray file) 
emitted from LEDs are used. For both models, spectrum files are also considered.
Figure~\ref{fig_spot} shows spot diagrams obtained 
at a distance 100 m from LED and free-form lens system on the rectangular 
detector whose size is 15 m $\times$ 15 m. 
Luminous flux of white LED is selected as 100 lm and radiant power of the 
IR LED is fixed to 1 W.
When using eulumdat files, as we expect, almost all of the rays fall on the 
detector and produce a very small light spot.
However, the use of ray files result in relatively wider spot sizes at the same distance.

In addition, the study is repeated for two more polymers as well;
PC (polycarbonate,  $(\text{C}_{16} \text{H}_{14} \text{O}_{3})_n)$ and 
PS (polystyrene, $(\text{C}_8 \text{H}_8)_n$).
The simulation performance for three materials when 
using ray and spectrum files are summarized in Table~\ref{tab_spot}.
Obviously, PMMA exhibits the best performance in terms of total optical power on the detector. 
This is because the transmittance of the PMMA in visible and near IR region 
is greater than the others. On the other hand, one can obtain a smaller 
RMS spot size on the detector when using PC as a lens material.

\begin{table}[H]
	\centering
	\renewcommand{\arraystretch}{1}
	\caption{Simulation results obtained by using ray and spectrum files of LEDs
		     for three polymeric materials. 
		     All values are calculated on 15 m$\times$ 15 m detector 
		     at 100 m from the lens. Total source power of the LEDs are 100 lm and 
		     1 W respectively. Total number of photon hits per total number of generated rays 
		     ($N=1\times 10^6$) on the detector are also given.}
	\begin{tabular}{lllll}
		\hline
		Quantity                &  LED    & PMMA     &   PC       &   PS        \\\hline
		RMS Spot diameter       &  White  & 2.5 m    &   2.4  m   &  2.6  m     \\
		Total luminous power    &  White  & 65.1 lm  &   57.7 lm  &  57.3 lm    \\
        Total hits / $N$        &  White  & 65.1 \%  &   57.5 \%  &  56.8 \%    \\
        \hline
        RMS Spot diameter       &   IR    & 3.0 m    &   2.3 m    &  2.8 m     \\
		Total radiant power     &   IR    & 0.45 W   &   0.36 W   &  0.35 W     \\
		Total hits / $N$        &   IR    & 45.5 \%  &   35.5 \%  &  34.7 \%    \\
		 \hline
	\end{tabular}
	\label{tab_spot}
\end{table}

\section{Manufacturing the Free-Form Lens and Building Projector}
\label{sec_manufacture}
\begin{figure}[H]
	\centering
	\includegraphics[width=0.49\textwidth]{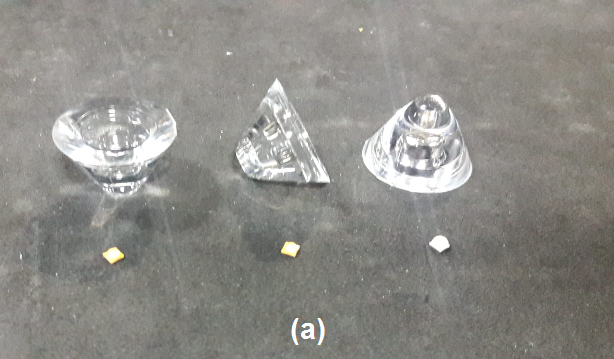}
	\includegraphics[width=0.49\textwidth]{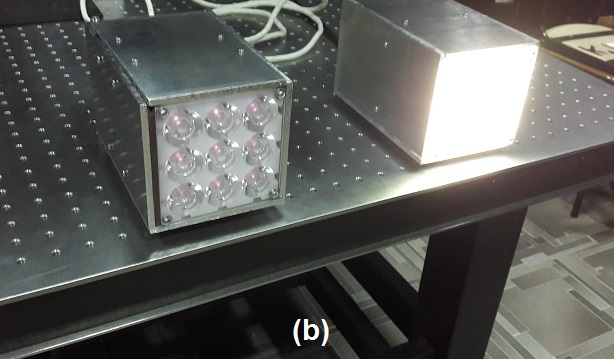}
	\caption{(a) Fabricated lenses made from PMMA and 
		LEDs used in the study. (b) Two projectors built 
		from the these lenses when low electrical powers are supplied.}
	\label{fig_solidlens}
\end{figure}

Manufacturers prefers mostly PMMA as a lens material since it has lower index of 
refraction and better optical transparency properties~\cite{Sultanova}. Therefore, 
a prototype of the proposed lens is manufactured 
by using PMMA via plastic injection molding method.
Figure~\ref{fig_solidlens}a shows the fabricated free-form lens 
for $p = q = 4$ mm, $h=20$ mm and 
$n=1.49$\footnote{This is the average refractive index 
  of PMMA for the wavelength range from 400 nm to 1000 nm.}.
In addition, two prototype projectors are built by combining nine lenses in the 
form of a 3x3 array as shown in Figure~\ref{fig_solidlens}b.

\section{Optical Performance of Projectors}
In order to test the optical performance of the projector, two measurements are carried out using 
IR LEDs (Osram SFH 4718A) and white LEDs (Osram LUW H9GP).
In the measurements, both projectors are supplied by an LED driver 
whose maximum electrical power is about 7.7 W ($V$ = 20-22 V and $I = 0.35$ A).

\begin{figure}[H]
	\centering
	\includegraphics[width=1.0\textwidth]{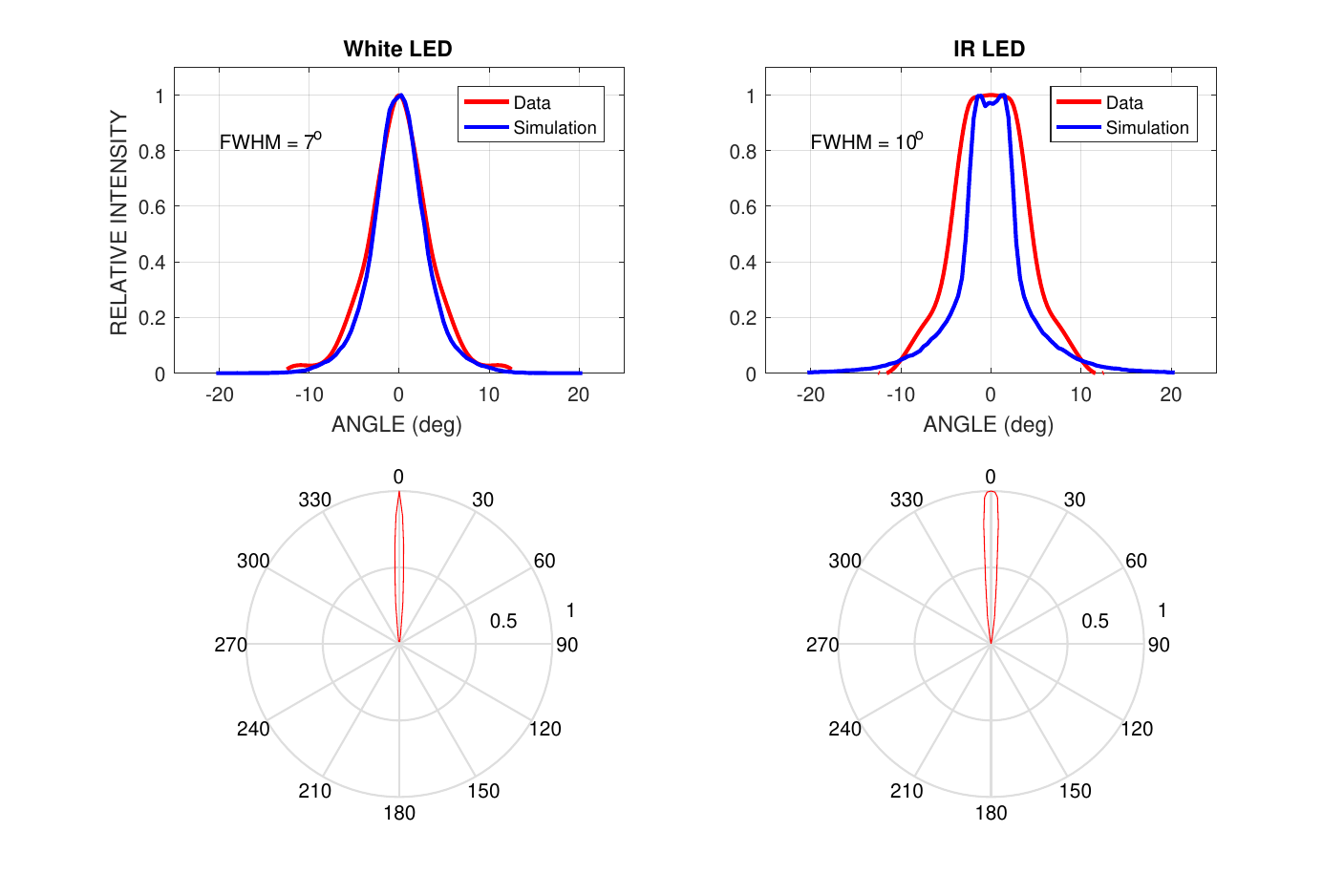}
	\caption{Comparison of goniophotometric measurements with simulations in both cartesian and polar coordinates.}
	\label{fig_angle}
\end{figure}

The simulation and experimental results for the angular light 
distributions are shown in Figure~\ref{fig_angle}.
It is obvious that the experimental data (solid red lines) are almost consistent with the 
simulations (solid blue lines). Full width half maximum (FWHM) values of the distributions
are approximately evaluated as $7.0^\circ$ for white light projector, 
and $9.3^\circ$ for IR projector. 
However, $FWHM \approx 6.1^\circ$ for IR projector in the simulation.
Apparently, both projectors can be used for long range (or narrow angle) illumination.

\begin{figure}[H]
	\centering
	\includegraphics[width=1.0\textwidth]{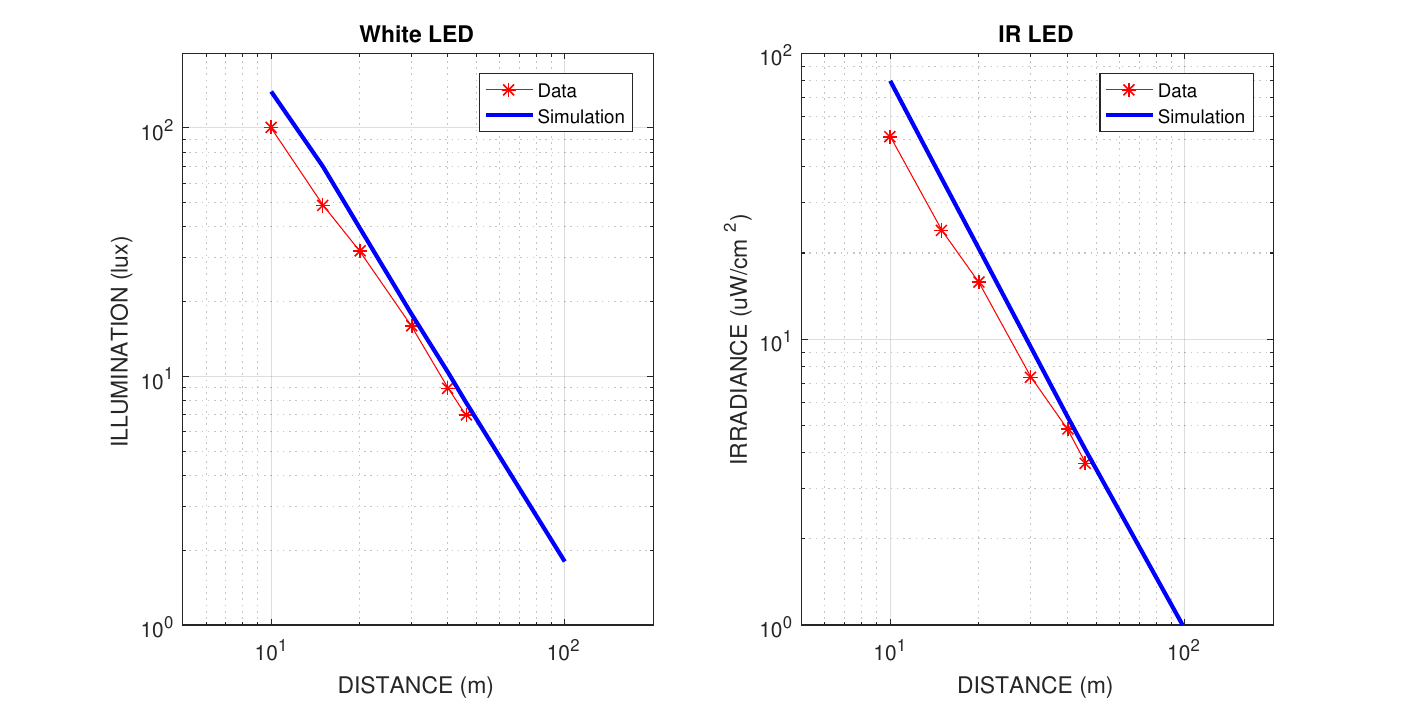}
	\caption{Simulation and measurement results of peak illuminance (left) and peak 
		     irradiance (right) as a function of distance from projectors.}
	\label{fig_distance}
\end{figure}

Second measurements are performed to test the dependence of light 
intensity as a function of distance.
In the IR region at around 850 nm, optical powers are measured by 
using Newport 843-R Powermeter.
The counterpart study is repeated for the white light projector.
In this case, illuminance data are acquired by means of Testo 435 Luxmeter.
Measurements are made at various distances as shown in Figure~\ref{fig_distance}.
The experimental data and simulation results are represented by red stars 
and by solid blue lines respectivley.
Measured values are less than that of simulation as we expect.
This is possible because the manufactured prototype projectors have some misalignments,
however, simulations represent the perfect geometry and alignment.

\begin{figure}[H]
	\centering
	\includegraphics[width=0.49\textwidth]{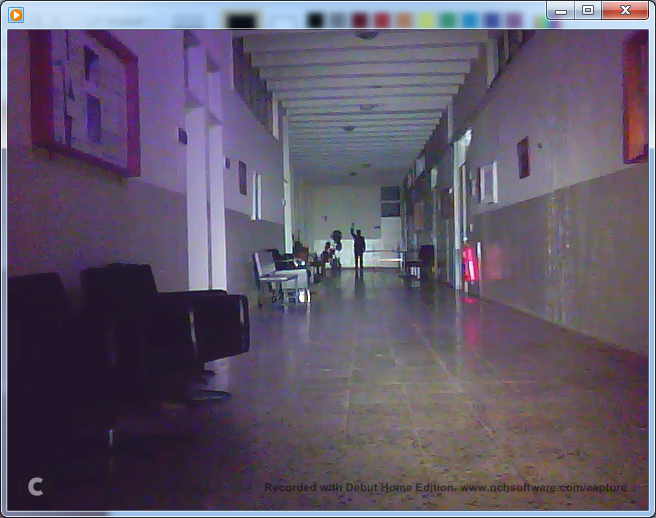}
	\includegraphics[width=0.49\textwidth]{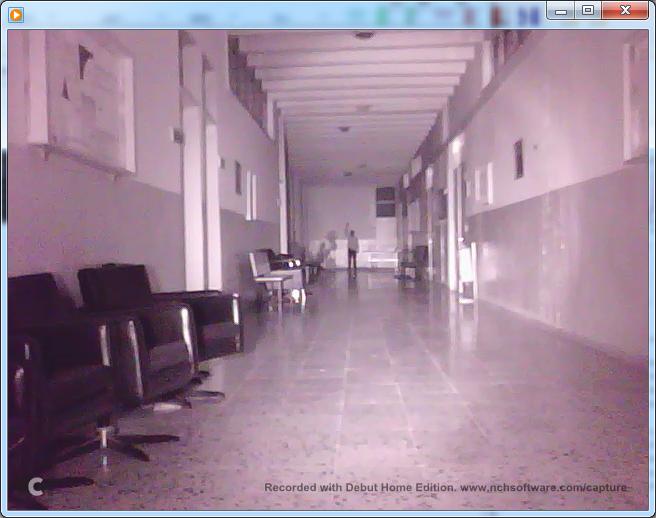}
	\caption{Two photographs of a 50 m long and completely dark corridor 
		taken by a webcam after illuminating by the projector with white LED (left) and
		with IR LED (right).}
	\label{fig_irwalk}
\end{figure}

Finally, the illumination performance of projectors
are shown in Figure~\ref{fig_irwalk}. 
Two photographs of a 50 m long and dark corridor are taken after
illumination\footnote{These photograph are taken by a webcam whose IR filter 
	is removed so that full IR waves can also be seen as in Figure~\ref{fig_irwalk} (right).}. 
The man at the end of corridor is seen clearly after illumination.
%
\section{Summary and Conclusion}
In this study, a prototype of a free-form lens which can be used for long range 
illumination is designed and manufactured. Surfaces of the free-form lens 
are extracted by equating the optical path length of the rays from a point-like light source. 
Ray tracing simulations are performed in Zemax OpticStudio 20.1
where IR and visible LEDs are selected as light sources and 
various optical polymers (PMMA, PC and PS) are concerned as lens material.

In manufacturing of the lens PMMA is selected since it exhibits the best performance 
in terms of optical power calculated on the detector according to simulation results. 
Manufacturing is fulfilled by well known technique called the plastic injection molding 
method. 
Two LED projector prototypes containing nine lenses are built for visible illumination 
and IR irradiation applications. The optical measurements of these projectors are 
found to be consistent with simulations.

Finally, authors suggest that a larger array of such free-form lenses 
can be used as a LED projector for the purpose of long range illumination
since its angular divergence is less than $10^\circ$ and it requires low electric power. 
These projectors may be useful in optical wireless communication, 
in illumination of car parks or stadiums, in headlights of vehicles and in houses or 
homesteads for the security purposes.
\section{Acknowledgement}
We wish to thank to the Scientific and Technological Research Council of Turkey
(T\"urkiye Bilimsel ve Teknolojik Ara\c{s}t{\i}rma Kurumu, T\"UBiTAK) for supporting us to manufacture the 
proposed free-form lens and the projector.

\end{document}